\documentclass[useAMS,usenatbib,usegraphicx]{mn2e}

\title[Optimizing interferometers for CMB B mode]
{Optimizing interferometer experiments for CMB B mode measurement}
\author[Jaiseung Kim]{Jaiseung Kim$^{1}$\thanks{E-mail:
jkim@physics.brown.edu}\\ 
$^{1}$ Department of Physics, Brown University, Box 1843, Providence, U.S.A.}

\begin{document}
\date{Accepted 2006 November 16. Received 2006 October 29; in original form 2006 July 26}
\pagerange{\pageref{firstpage}--\pageref{lastpage}} \pubyear{2006}
\maketitle
\label{firstpage}
\begin{abstract}
The sensitivity of interferometers with linear polarizers to the CMB E and B mode are variant under the rotation of the polarizer frame, while interferometer with circular polarizers
are equally sensitive to E and B mode. We present analytically and numerically that the diagonal elements of window functions for CMB E/B power spectra are maximized in interferometric measurement of linear polarization, when the polarizer frame is in certain rotation from the associated baseline. We also present the simulated observation to show that 
the 1$\sigma$ errors on E/B mode power spectrum estimation are variant under the polarizer frame rotation in the case of linear polarizers, while they are invariant in the case of circular polarizers.
Simulation of the configuration similar to the DASI shows that minimum 1$\sigma$ error on B mode in interferometer measurement with linear polarizers is 26 per cent of that in interferometric measurement with circular polarizers. The simulation also shows that the E/B mixing in interferometer measurement with linear polarizers can be as low as 23 per cent of that in interferometric measurement with circular polarizers.
It is not always possible to physically align the polarizer frame with all the associated baselines in the case of an interferometer array (N$>$2). There exist certain linear combinations of visibilities, which are equivalent to visibilities of the optimal polarizer frame rotation. We present the linear combinations, which enables B mode optimization for an interferometer array (N$>$2). 
\end{abstract}

 
\section{Introduction}
The Cosmic Microwave Background (CMB) is expected to be linearly polarized by Thomson scattering at the last scattering surface and after re-ionization.
The detection of the CMB polarization has been reported by the Degree Angular Scale Interferometer (DASI) \citep{DASI:data} and recently by the Wilkinson Microwave Anisotropy Probe (WMAP) satellite \citep{WMAP:polarization}.
Measurements of the CMB temperature anisotropy with interferometers are made in the Very Small Array (VSA), the Cosmic Background Imager (CBI) and
many other experiments. The CMB polarization measurements with interferometers are on-going and planned in the experiments such as DASI, CBI and the Millimeter-wave Bolometric Interferometer (MBI) \citep{Tucker:Bolometric_interferometry,SPIE:MBI}. With many desirable features of an interferometer, interferometers are more and more employed in CMB polarization experiments.

The CMB polarization can be decomposed into gradient-like E mode and curl-like B mode \citep{Seljak-Zaldarriaga:Polarization}.
B mode polarization is not induced by scalar density perturbation but only tensor perturbation, while E mode polarization is induced by both
\citep{Seljak:signature_gravity_wave}. 
Since tensor-to-scalar ratio is much smaller than unity in most inflationary models, B mode polarization is expected to be much smaller than 
E mode polarization. 
Though there is complication by gravitational lensing \citep{CMB_Lensing:full-sky},
measurement of B mode polarization makes it possible probing the Universe on the energy scale at inflationary period \citep{Modern_Cosmology}. 

The 1$\sigma$ error on the parameter estimation can be forecast from the Fisher matrix \citep{Modern_Cosmology}.
It will be shown that in interferometric measurement of Stokes parameter $Q$ or $U$, 
the 1$\sigma$ error on E and B power spectra estimation varies with rotation of its polarizer frame. We will show that certain rotation of the polarizer frame from the associated baseline minimize the 1$\sigma$ error on either E or B mode power spectra estimation.  
For a feedhorn array (N$>$2), it is not always possible to realize the specific rotation of the polarizer frame from all the associated baselines.
We will show that forming certain linear combinations of polarimetric visibilities is identical with physically rotating the polarizer frame to the optimal orientation, thereby enabling the B mode measurement optimization for a feedhorn array (N$>$2).  

This paper is organized as follows.
We discuss Stokes parameters in \S 2. The formalism for interferometric CMB polarization measurement on a spherical sky is presented \S 3. 
In \S 4, with flat sky approximation, we show that interferometric measurement of Stokes parameter Q or U can be configured to suppressing E mode, leading to smaller leakage between E and B mode.
In \S 5, we discuss the variation of 1$\sigma$ error on power spectra estimation and show that B mode sensitivity is maximized, when the polarizer frame is in certain rotation from the baseline.
In \S 6, we show that there exists certain linear combinations of visibilities, which enables the B mode optimization for a feedhorn array (N$>$2). 
In \S 7, the summary and conclusion are given. 
In appendix, we show analytically that the diagonal element of window functions are maximized, when the polarizer frame is rotated from the baseline by certain angles.
\section{STOKES PARAMETERS}
\label{Stokes}
There are Stokes parameters, which describe the state of polarization \citep{Kraus:Radio_Astronomy},
which are measured in reference to $(\mathbf {\hat e_\theta},\mathbf {\hat e_\phi})$ \citep{Seljak-Zaldarriaga:Polarization}. $\hat e_\theta$ and $\hat e_\phi$  are the unit vectors of the spherical coordinate system and given by \citep{Arfken}
\begin{eqnarray*}
\mathbf {\hat e_{\theta}}&=&\mathbf {\hat i}\;\cos\theta\cos\phi+\mathbf {\hat j}\;\cos\theta\sin\phi-\mathbf {\hat k}\;\sin\theta,\\
\mathbf {\hat e_{\phi}}&=&-\mathbf {\hat i}\;\sin\phi+\mathbf {\hat j}\;\cos\phi.
\end{eqnarray*}

Since Thomson scattering does not generate circular polarization but linear polarization in early universe, the phase difference between $E_\theta$ and $E_\phi$ is zero and circular polarization state $V$ is redundant in the study on the CMB polarization.
Stokes parameters Q and U are as follows:
\begin{eqnarray} 
\label{Q} Q&=&\left\langle E_\theta^2-E_\phi^2\right\rangle,\\
\label{U} U&=&\left\langle 2E_\theta\,E_\phi\right\rangle,
\end{eqnarray}
where $\left\langle \ldots\right\rangle$ indicates time average.
$Q$ and $U$ transform under rotation of an angle $\psi$ on the plane perpendicular to direction $\mathbf {\hat n}$ as  
\begin{eqnarray}
\label{Q'} Q'(\mathbf {\hat n})&=&Q(\mathbf {\hat n})\cos2\psi+U(\mathbf {\hat n})\sin2\psi,\\
\label{U'} U'(\mathbf {\hat n})&=&-Q(\mathbf {\hat n})\sin2\psi+U(\mathbf {\hat n})\cos2\psi,
\end{eqnarray}
with which the following quantities can be constructed \citep{Seljak-Zaldarriaga:Polarization}:
\begin{eqnarray}
\label{Q'U'} Q^\prime(\mathbf {\hat n})\pm i U^\prime(\mathbf {\hat n})=e^{\mp 2i\psi}(Q(\mathbf {\hat n})\pm i U(\mathbf {\hat n})).
\end{eqnarray}

For all-sky analysis, $Q$ and $U$ are expanded in terms of spin $\pm2$ spherical harmonics \citep{Seljak-Zaldarriaga:Polarization} as follows:
\begin{eqnarray}
\label{Q_lm+iU_lm} Q(\hat {\mathbf n})+i U(\hat {\mathbf n})&=&\sum_{l,m} -(a_{E,lm}+i\,a_{B,lm})\;{}_2Y_{lm}(\hat {\mathbf n}),\\
\label{Q_lm-iU_lm} Q(\hat {\mathbf n})-i U(\hat {\mathbf n})&=&\sum_{l,m} -(a_{E,lm}-i\,a_{B,lm})\;{}_{-2}Y_{lm}(\hat {\mathbf n}),
\end{eqnarray}
\section{Interferometric Measurement}
The discussion in this section is for an ideal interferometer.
An interferometer measures time-averaged correlation of two electric field from a pair of identical apertures positioned at $\mathbf r_1$ and at $\mathbf r_2$. 
The separation, $\mathbf B=\mathbf r_1-\mathbf r_2$, of two apertures is called the `baseline' and the measured correlation is called `visibility' \citep{Lawson:Interferometry}.
Depending on the instrumental configuration, visibilities
are associated with $\langle E_x^2-E_y^2\rangle$, $\langle 2 E_x E_y\rangle$ and $\langle E_x^2-E_y^2\pm \imath\,2 E_x E_y\rangle$ respectively, where $\hat x$ and $\hat y$ are axes of the polarizer frame.
As discussed in \S \ref{Stokes}, Stokes parameters at angular coordinate ($\theta$,$\phi$) are defined in respect to two basis vectors $\mathbf {\hat e_\theta}$ and $\mathbf {\hat e_\phi}$.
Consider the polarization observation, whose antenna pointing is in the direction of angular coordinate ($\theta_A$,$\phi_A$). The polarizers and baselines are assumed to be on the aperture plane. Then, the global frame coincides with the polarizer frame after Euler rotation ($\phi_A$, $\theta_A$, $\psi$) on the global frame, where $\psi$ is the rotation around the axis in the direction of antenna pointing.
Most of interferometer experiments for the CMB observation employ feedhorns for beam collection.
After passing through the feedhorn system, an incoming off-axis ray becomes on-axis ray.
Then, the basis vectors $\mathbf {\hat e_\theta}$ and $\mathbf {\hat e_\phi}$ of the ray after the feedhorn system are related to the basis vectors $\mathbf {\hat e_x}$ and $\mathbf {\hat e_y}$ 
of the polarizer frame as follows:
\begin{eqnarray}
\label{azimuthal_rotation}
\mathbf {\hat e_{x}}+\imath\,\mathbf {\hat e_{y}}=
e^{-\imath\psi}(\mathbf {\hat e_{\theta_A}}+\imath\,\mathbf {\hat e_{\phi_A}})
=e^{\imath(\Phi-\psi)}(\mathbf {\hat e_\theta}+\imath\,\mathbf {\hat e_\phi}),
\end{eqnarray}
where $\Phi$ is given by 
\begin{eqnarray*}
\label{Phi}\Phi&=&\tan^{-1}\left[\frac{\sin\theta\sin(\phi-\phi_A)}{\sin\theta\cos\theta_A\cos(\phi-\phi_A)-\cos\theta\sin\theta_A}\right]\\
&&+\tan^{-1}\left[\frac{\sin\theta_A\sin(\phi-\phi_A)}{-\sin\theta\cos\theta_A+\cos\theta\sin\theta_A\cos(\phi-\phi_A)} \right].
\end{eqnarray*}
Refer to Appendix \ref{relation} for the details on the derivation of $\Phi$.
With Eq. \ref{azimuthal_rotation}, we can easily show that
\begin{eqnarray*}
\langle E_x^2- E_y^2\rangle+\imath\langle 2E_x\,E_y\rangle=e^{-\imath(2\psi-2\Phi)}
(\langle E_\theta^2- E_\phi^2\rangle+\imath\langle 2E_\theta\,E_\phi\rangle).
\end{eqnarray*}
With the employment of linear polarizers, the visibilities associated with $\langle E_x^2-E_y^2\rangle$ or $\langle 2 E_x E_y\rangle$ are as follows:
\begin{eqnarray}
\label{V_Q_spherical}
V_{Q^\prime}&=&f(\nu)\int \mathrm d \Omega A(\mathbf {\hat n}-\hat {\mathbf n}_A)\\
&&\times\mathrm{Re}\,[e^{-i(2\psi-2\Phi(\mathbf {\hat n}))}
(Q(\mathbf {\hat n})+i U(\mathbf {\hat n}))]\,e^{i\,2\pi\mathbf u\cdot \mathbf {\hat n}},\nonumber\\
\label{V_U_spherical}
V_{U^\prime}&=&f(\nu)\int \mathrm d \Omega A(\mathbf {\hat n}-\hat {\mathbf n}_A)\\
&&\times\mathrm{Im}\,[e^{-i(2\psi-2\Phi(\mathbf {\hat n}))}
(Q(\mathbf {\hat n})+i U(\mathbf {\hat n}))]\,e^{i\,2\pi\mathbf u\cdot \mathbf {\hat n}},\nonumber
\end{eqnarray}
where $\hat {\mathbf n}_A$ is the  unit vector in the direction of antenna pointing and
$f(\nu)$ is the frequency spectrum of the CMB polarization. \footnote{$f(\nu)=\left.\frac {\partial B(\nu,T)} {\partial T}\right|_{T=T_0}$, where $B(\nu,T)$ is the Plank function and
$T_0$ is the CMB monopole temperature.}
With the employment of circular polarizers, the visibilities associated with $\langle E_x^2-E_y^2\pm i\,2 E_x E_y\rangle$ are as follows: 
\begin{eqnarray}
\label{V_RL_spherical}
V_{RL}&=&f(\nu)\int \mathrm d \Omega A(\mathbf {\hat n}-\hat {\mathbf n}_A),\\ 
&&\times [Q(\mathbf {\hat n})+i U(\mathbf {\hat n})]e^{i(2\pi\mathbf u\cdot \mathbf {\hat n}-2\psi+2\Phi(\mathbf {\hat n}))},\nonumber\\
\label{V_LR_spherical}
V_{LR}&=&f(\nu)\int \mathrm d \Omega A(\mathbf {\hat n}-\hat {\mathbf n}_A),\\
&&\times [Q(\mathbf {\hat n})-i U(\mathbf {\hat n})]e^{i(2\pi\mathbf u\cdot \mathbf {\hat n}+2\psi-2\Phi(\mathbf {\hat n}))},\nonumber
\end{eqnarray}
where $R$ and $L$ stand for right/left circular polarizers.

With Eq. \ref{Q_lm+iU_lm} and \ref{Q_lm-iU_lm}, visibilities are expressed in terms of E/B mode  as follows:
\begin{eqnarray}
\label{V_Q_EB_spherical}
V_{Q^\prime}&=&-f(\nu)\int \mathrm d \Omega A(\mathbf {\hat n}-\hat {\mathbf n}_A)\,e^{i(2\pi\mathbf u\cdot \mathbf {\hat n})}\\ &&\times\mathrm{Re}\left[e^{-i(2\psi-2\Phi(\mathbf {\hat n}))}\,(a_{E,lm}+i\,a_{B,lm})\;{}_2Y_{lm}\right],\nonumber\\
\label{V_U_EB_spherical}
V_{U^\prime}&=&-f(\nu)\int \mathrm d \Omega A(\mathbf {\hat n}-\hat {\mathbf n}_A)\,e^{i(2\pi\mathbf u\cdot \mathbf {\hat n})}\\ &&\times\mathrm{Im}\left[e^{-i(2\psi-2\Phi(\mathbf {\hat n}))}\,(a_{E,lm}+i\,a_{B,lm})\;{}_2Y_{lm}\right],\nonumber\\
\label{V_RL_EB_spherical}
V_{RL}&=&-f(\nu)\int \mathrm d \Omega A(\mathbf {\hat n}-\hat {\mathbf n}_A)\\ 
&&\times (a_{E,lm}+i\,a_{B,lm})\;{}_2Y_{lm}e^{i(2\pi\mathbf u\cdot \mathbf {\hat n}-2\psi+2\Phi(\mathbf {\hat n}))},\nonumber\\
\label{V_LR_EB_spherical}
V_{LR}&=&-f(\nu)\int \mathrm d \Omega A(\mathbf {\hat n}-\hat {\mathbf n}_A)\\
&&\times (a_{E,lm}-i\,a_{B,lm})\;{}_{-2}Y_{lm}e^{i(2\pi\mathbf u\cdot \mathbf {\hat n}+2\psi-2\Phi(\mathbf {\hat n}))}.\nonumber
\end{eqnarray}
  
\section{Reducing leakage between E and B mode}
\label{V_flat_sky}
For the observation of small patch of sky, flat sky approximation in small angle limit can be used. In flat sky approximation, Eq. \ref{V_Q_spherical}, \ref{V_U_spherical}, \ref{V_RL_spherical} and \ref{V_LR_spherical} are as follows:
\begin{eqnarray*}
V_{Q^\prime}&=&f(\nu)\int \mathrm d \mathbf x^2 A(\mathbf x)\\
&&\times\mathrm{Re} \left[e^{-2i\,\psi}\{Q(\mathbf x)+i\,U(\mathbf x)\}\right]e^{i 2\pi \mathbf u \cdot \mathbf x},\\
V_{U^\prime}&=&f(\nu)\int \mathrm d \mathbf x^2 A(\mathbf x)\\
&&\times\mathrm{Im} \left[e^{-2i\,\psi}\{Q(\mathbf x)+i\,U(\mathbf x)\}\right]e^{i 2\pi \mathbf u \cdot \mathbf x},\\
V_{RL}&=&f(\nu)\int \mathrm d \mathbf x^2 A(\mathbf x) [Q(\mathbf x)+i U(\mathbf x)]e^{i(2\pi\mathbf u\cdot \mathbf x-2\psi)},\\
V_{LR}&=&f(\nu)\int \mathrm d \mathbf x^2 A(\mathbf x) [Q(\mathbf x)-i U(\mathbf x)]e^{i(2\pi\mathbf u\cdot \mathbf x+2\psi)},
\end{eqnarray*}
where $A(\mathbf x)$ is the function of beam power pattern.
Since visibilities are the convolution of the Fourier transform of a beam function
with the Fourier transform of $Q$ and $U$\citep{Hobson:maximum_likelihood}, they can be written as follows:
\begin{eqnarray*}
V_{Q^\prime}&=&f(\nu) \int \mathrm d^2 \mathbf u^\prime\tilde A(\mathbf u-\mathbf u^\prime)\\
&&\times[\cos(2\,\psi)\tilde Q(\mathbf u^\prime)+\sin(2\,\psi)\,\tilde U(\mathbf u^\prime)],\\
V_{U^\prime}&=&f(\nu) \int \mathrm d^2 \mathbf u^\prime \tilde A(\mathbf u-\mathbf u^\prime)\\
&&\times[-\sin(2\,\psi)\tilde Q(\mathbf u^\prime)+\cos(2\,\psi)\,\tilde U(\mathbf u^\prime)],\\
V_{RL}&=&f(\nu) \int \mathrm d^2 \mathbf u^\prime \tilde A(\mathbf u-\mathbf u^\prime)\;[e^{-i2\,\psi}(\tilde Q(\mathbf u^\prime)+i\,\tilde U(\mathbf u^\prime))],\\
V_{LR}&=&f(\nu) \int \mathrm d^2 \mathbf u^\prime \tilde A(\mathbf u-\mathbf u^\prime)\;[e^{i2\,\psi}(\tilde Q(\mathbf u^\prime)-i\,\tilde U(\mathbf u^\prime))],\\
\end{eqnarray*}
where tilde $\tilde {}$ indicates Fourier transform. 
In the flat sky approximation in small angle limit, Stokes parameter Q and U can be decomposed as follows \citep{Seljak-Zaldarriaga:Polarization}: 
\begin{eqnarray}
\label{Q_EB_flat}
\tilde Q(\mathbf u)&=&\cos(2\phi_{\mathbf u})\,\tilde E(\mathbf u)-\sin(2\phi_{\mathbf u})\,\tilde B(\mathbf u)),\\
\label{U_EB_flat}
\tilde U(\mathbf u)&=&\sin(2\phi_{\mathbf u})\,\tilde E(\mathbf u)+\cos(2\phi_{\mathbf u})\,\tilde B(\mathbf u)),
\end{eqnarray}
where $\phi_{\mathbf u}$ is the direction angle of a vector $\mathbf u$. 
With Eq. \ref{Q_EB_flat} and \ref{U_EB_flat}, visibilities are expressed in terms of $E$ and $B$ mode as follows:
\begin{eqnarray}
\label{V_Q_EB_flat} V_{Q^\prime}&=&f(\nu) \int \mathrm d^2 \mathbf u^\prime\tilde A(\mathbf u-\mathbf u^\prime)\\
&&\times[\cos(2(\psi-\phi_{\mathbf u^\prime}))\tilde E(\mathbf u^\prime)+\sin(2(\psi-\phi_{\mathbf u^\prime}))\,\tilde B(\mathbf u^\prime)]\nonumber,\\
\label{V_U_EB_flat} V_{U^\prime}&=&f(\nu) \int \mathrm d^2 \mathbf u^\prime\tilde A(\mathbf u-\mathbf u^\prime)\\
&&\times[-\sin(2(\psi-\phi_{\mathbf u^\prime}))\tilde E(\mathbf u^\prime)+\cos(2(\psi-\phi_{\mathbf u^\prime}))\,\tilde B(\mathbf u^\prime)]\nonumber,\\
\label{V_RL_EB_flat} V_{RL}&=&f(\nu) \int \mathrm d^2 \mathbf u^\prime\tilde A(\mathbf u-\mathbf u^\prime)\\
&&\times\left[e^{-i2\,(\psi-\phi_{\mathbf u^\prime})}(\tilde E(\mathbf u^\prime)+i\,\tilde B(\mathbf u^\prime))\right]\nonumber,\\
\label{V_LR_EB_flat}V_{LR}&=&f(\nu) \int \mathrm d^2 \mathbf u^\prime\tilde A(\mathbf u-\mathbf u^\prime)\\
&&\times\left[e^{i2\,(\psi-\phi_{\mathbf u^\prime})}(\tilde E(\mathbf u^\prime)-i\,\tilde B(\mathbf u^\prime))\right]\nonumber.
\end{eqnarray}
A Gaussian beam is a good approximation to many CMB experiments and the Fourier transform of a Gaussian beam is $\exp[-\frac{|\mathbf u-\mathbf u^\prime|^2\sigma^2}{2}]$, where
$\sigma=0.4245\;\mathrm{FWHM}$.
With such a beam, biggest contribution comes from $\mathbf u^\prime=\mathbf u$ in the integration over $\mathbf u^\prime$. We can approximate visibilities as follows:
\begin{eqnarray}
\label{V_Q_FOV}
V_{Q^\prime}&\approx&f(\nu)\\
&&\times[\cos(2(\psi-\phi_{\mathbf u}))\tilde E(\mathbf u)+\sin(2(\psi-\phi_{\mathbf u}))\,\tilde B(\mathbf u)],\nonumber\\
\label{V_U_FOV}
V_{U^\prime}&\approx&f(\nu)\\
&&\times[-\sin(2(\psi-\phi_{\mathbf u}))\tilde E(\mathbf u)+\cos(2(\psi-\phi_{\mathbf u}))\,\tilde B(\mathbf u)],\nonumber\\
\label{V_RL_FOV}
V_{RL}&\approx&f(\nu) \left[e^{-i2\,(\psi-\phi_{\mathbf u})}(\tilde E(\mathbf u)+i\,\tilde B(\mathbf u))\right],\\
\label{V_LR_FOV}
V_{LR}&\approx&f(\nu) \left[e^{i2\,(\psi-\phi_{\mathbf u})}(\tilde E(\mathbf u)-i\,\tilde B(\mathbf u))\right].
\end{eqnarray}
As seen in Eq. \ref{V_RL_FOV} and \ref{V_LR_FOV}, the measurement of $V_{RL}$ and $V_{LR}$ measures E and B mode equally, independent of the polarizer rotation.
From Eq. \ref{V_Q_FOV} and \ref{V_U_FOV}, it is seen that 
$V_{Q^\prime}$ and $V_{U^\prime}$ gets unequal contribution from E and B mode, when the baseline and the polarizer frame are aligned as follows:
\begin{eqnarray*}
V_{Q^\prime}&\approx&\left\{\begin{array}{r@{\quad:\quad}l}
f(\nu)\,\tilde E(\mathbf u)&\psi=\phi_{\mathbf u}\\
f(\nu)\,\tilde B(\mathbf u)&\psi=\phi_{\mathbf u}+\pi/4\end{array}\right.\\
V_{U^\prime}&\approx&\left\{\begin{array}{r@{\quad:\quad}l}
f(\nu)\,\tilde B(\mathbf u)&\psi=\phi_{\mathbf u}\\
-f(\nu)\,\tilde E(\mathbf u)&\psi=\phi_{\mathbf u}+\pi/4\end{array}\right.\\
\end{eqnarray*}
Physically, $\psi=\phi_{\mathbf u}$ corresponds to aligning $x$ axis of the polarizer frame with the baseline, and $\psi=\phi_{\mathbf u}+\pi/4$ is rotating 
the $x$ axis of the polarizer frame from the baseline by $45^\circ$ on the aperture plane.
Compared with those of $V_{RL}$ and $V_{RL}$ in those configuration, either E or B mode in $V_{Q^\prime}$ and $V_{U^\prime}$ is suppressed while the other mode is intact. 
The complete separation of E mode from B mode is not possible unless a full-sky map is made with infinite angular resolution \citep{Bunn:EB-Separation}. 
The leak from E mode into much weaker B mode causes serious problem. 
We can reduce E mode leak into B mode by measuring $V_{Q^\prime}$ with $\psi=\phi_{\mathbf u}+\pi/4$ and  $V_{U^\prime}$ with $\psi=\phi_{\mathbf u}$, in which E mode contribution is suppressed. 

\begin{figure}
\begin{center}
\includegraphics[scale=.5]{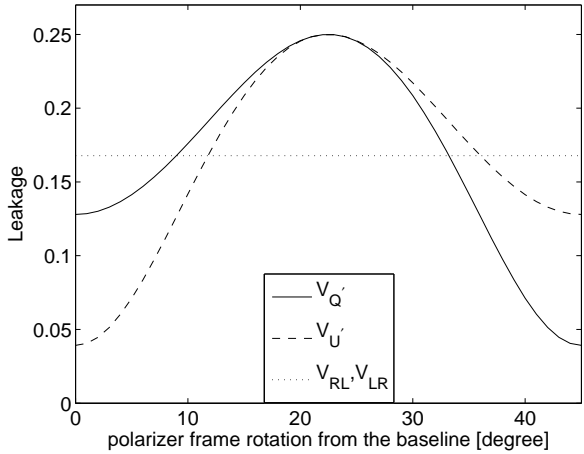}
\end{center}
\caption{$L_{EB}$ and $L_{BE}$ for various $\psi-\phi_{\mathbf u}$}
\label{leakage}
\end{figure}
We have computed the $2\times2$ leakage matrix \citep{Tegmark:measuring_polarization}
for the simulated observation in \S \ref{simulation}. $L_{EB}$ indicates the B mode leakage into E mode measurement and $L_{BE}$ indicates the E mode leakage into B mode measurement. If $L_{EB}=L_{BE}=0$, there is no leakage at all \citep{Tegmark:measuring_polarization}. 
$L_{EB}$ and $L_{BE}$ are shown for various $\psi-\phi_{\mathbf u}$ in Fig. \ref{leakage}, where $L_{EB}$ and $L_{BE}$ are dimensionless, and visually identical. 
It shows that the leakage in $V_{Q^\prime}$ ($V_{U^\prime}$) is smallest at $\psi=\phi_{\mathbf u}+\pi/4$ ($\psi=\phi_{\mathbf u}$). 
It is seen that with the choice of $\psi-\phi_{\mathbf u}$,
the leakage of $V_{Q^\prime}$ and $V_{U^\prime}$  can be as low as 23 per cent of that of $V_{RL}$ and $V_{LR}$.
 
\section{variation of 1$\sigma$ error}
\label{simulation}
It has become standard to estimate band power spectra \citep{Knox:Window} by maximum likelihood method  from CMB interferometric observations \citep{Hobson:likelihood_interferometer}.
With the Gaussianity of polarimetric visibility and noise, likelihood function is given by
\[\mathcal{L}=\frac{1}{(2\pi)^{\frac{N}{2}}{\left|\mathbf S+\mathbf N\right|}^{\frac{1}{2}}}\exp[-\frac{1}{2}\Delta {(\mathbf S+\mathbf N)}^{-1}\Delta],\]
where $\Delta$ is data, $\mathbf S$ is signal covariance matrix and 
$\mathbf N$ is noise covariance matrix.
By exploring parameter space to maximize likelihood, we can estimate parameters within certain error limits. 
The diagonal element of $\partial\mathbf S/\partial C_{l}$, where $C_{l}$ is the angular power spectrum, shows the sensitivity of the experiment over multipoles. 
\begin{figure}
\begin{center}
\includegraphics[scale=.5]{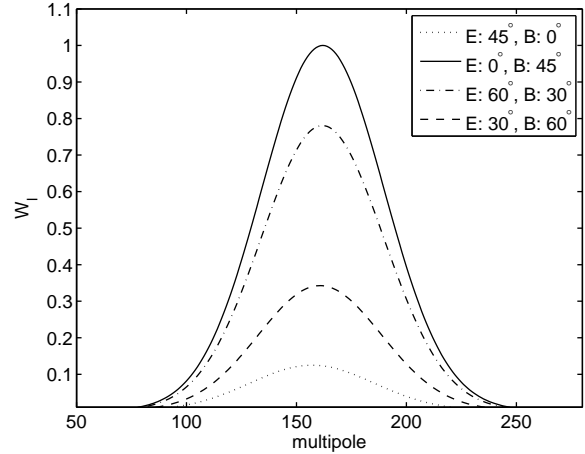}
\end{center}
\caption{Window functions in $V_{Q^\prime}$ measurement with $3.4^\circ$ FWHM}
\label{Wl_Q_3_plot}
\end{figure}
\begin{figure}
\begin{center}
\includegraphics[scale=.5]{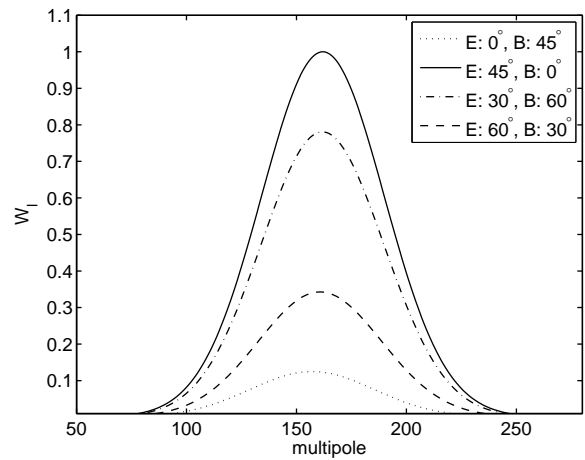}
\end{center}
\caption{Window functions in $V_{U^\prime}$ measurement with $3.4^\circ$ FWHM}
\label{Wl_U_3_plot}
\end{figure}
Through this paper, a window function means ${W}_{l}=\partial\mathbf S/\partial C_{l}$. In Appendix \ref{window_function}, we have shown analytically that the diagonal ${W}^{BB}_{l}$ of $V_{Q^\prime}$ is maximized at $\psi=\phi_{\mathbf u}\pm\pi/4$ while that of $V_{U^\prime}$ is maximized at $\psi=\phi_{\mathbf u},\;\phi_{\mathbf u}\pm\pi/2$. 
We have numerically computed ${W}^{EE}_{l}$ and ${W}^{BB}_{l}$ for an interferometer of a 25 cm baseline with 30 $\sim$ 31 GHz signal frequency range and $3.4^\circ$ Full Width at Half Maximum (FWHM) beam.  They are shown in Fig. \ref{Wl_Q_3_plot} and \ref{Wl_U_3_plot}, 
and agree with the analytical result in Appendix \ref{window_function}.
The window functions are normalized so that the peak value of the highest window function is a unit value. As also shown in Fig. \ref{Wl_Q_3_plot} and \ref{Wl_U_3_plot}, an interferometer is sensitive to the multipole range $l\approx 2\pi u\pm \Delta l/2$, where $u$ is a baselinelength divided by wavelength and $\Delta l$ is FWHM of the window function (for a circular Gaussian beam, $\Delta l=4\sqrt{2}\ln2/\theta_{\mathrm {FWHM}}$). In maximum likelihood estimation, it is usual to estimate the band powers \citep{Knox:Window}, which are assumed to be flat over some multipole range.
In an interferometer experiment, E and B mode band power, $\lambda_{EE}$ and $\lambda_{BB}$
are assumed to be flat over the multipole range the interferometer is sensitive to.

\begin{figure}
\begin{center}
\includegraphics[scale=.5]{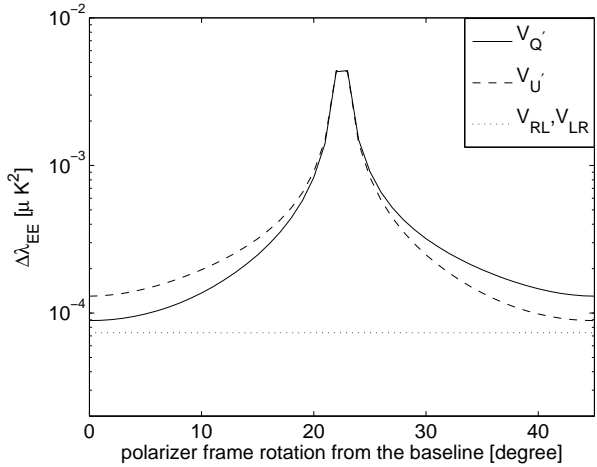}
\end{center}
\caption{$\Delta \lambda_{EE}$ for various $\psi-\phi_{\mathbf u}$}
\label{error_E}
\end{figure}

\begin{figure}
\begin{center}
\includegraphics[scale=.5]{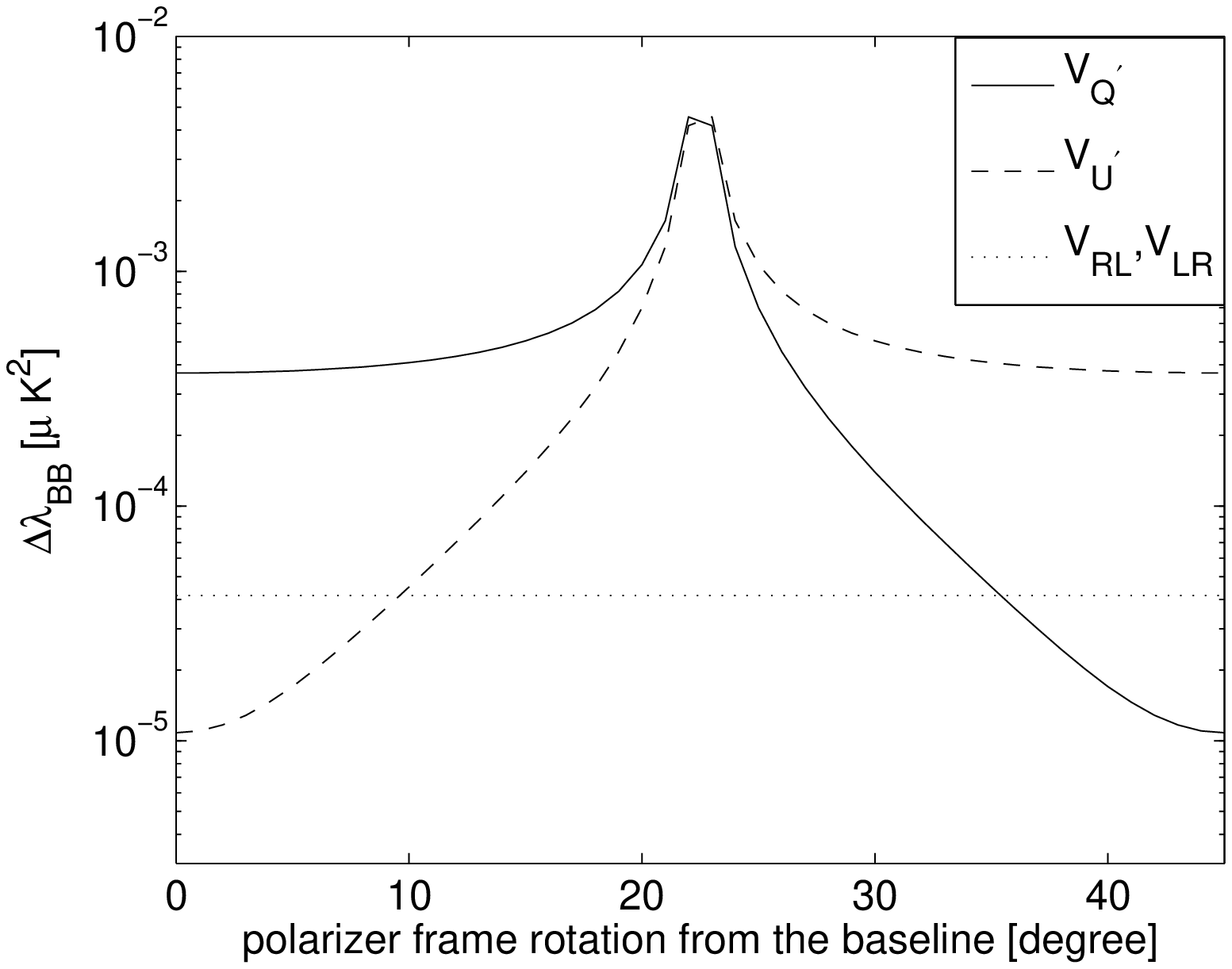}
\end{center}
\caption{$\Delta \lambda_{BB}$ for various $\psi-\phi_{\mathbf u}$}
\label{error_B}
\end{figure}

The minimum possible variance on the parameter estimation can be forecast from the Fisher matrix \citep{Modern_Cosmology},
which is defined as
\begin{eqnarray}
\label{Fisher_matrix}
\mathcal F_{ij}&=&\langle-\frac{\partial^2(\ln\mathcal{L})}{\partial \lambda_i\partial \lambda_j}\rangle\\
&=&\frac{1}{2} \mathrm{Tr}[\frac{\partial (\mathbf S+\mathbf N)}{\partial \lambda_i} (\mathbf S+\mathbf N)^{-1}\frac{\partial (\mathbf S+\mathbf N)}{\partial \lambda_j}(\mathbf S+\mathbf N)^{-1}]\nonumber.
\end{eqnarray}
Evaluated at the maximum of the likelihood, the square root of diagonal element of the inverse Fisher matrix yields the marginalized $1\sigma$ error on the parameter estimation.
As shown analytically in appendix \ref{window_function} and numerically in Fig. \ref{Wl_Q_3_plot} and \ref{Wl_U_3_plot}, diagonal windows functions of $V_{Q^\prime}$ and
$V_{U^\prime}$ are maximized at certain $\psi-\phi_{\mathbf u}$.
Therefore, $\Delta \lambda_{EE}$ and $\Delta \lambda_{BB}$ are expected to be smallest with certain $\psi-\phi_{\mathbf u}$.
We have numerically computed $\Delta \lambda_{EE}$ and $\Delta \lambda_{BB}$ from simulated experiments for various $\psi-\phi_{\mathbf u}$, which are shown in Fig. \ref{error_E} and \ref{error_B}. In the simulation, we used the Code for Anisotropies in the Microwave Background (CAMB)\citep{CAMB} to compute the power spectra of $\Lambda CDM$ with the tensor-to-scalar ratio ($r=0.3$). 
The baseline length 25 [cm] is assumed with the signal frequency range 30$\sim$31GHz with 1GHz bandwidth so that the interferometer probes the multipole range of roughly E1/B1 band of DASI\citep{DASI:data}.
The noise covariance matrix is assumed to be diagonal \citep{White:interferometer}
 and have a uniform value, for which we assumed the sensitivity of DASI: $60\times .1^2\pi\;\mathrm{Jy\,s^{1/2}\,m^2}$\citep{DASI:I}.
We assumed the probe of three fields, three hundred sixty five days integration time for each field and simultaneous thirty six orientations for each baseline length. The equatorial coordinates of the assumed three fields are ($80^\circ$, $0^\circ$), ($80^\circ$, $120^\circ$) and ($80^\circ$, $240^\circ$).
Fig. \ref{error_E} and \ref{error_B} shows $\Delta \lambda_{EE}$ 
and $\Delta \lambda_{BB}$ for various $\psi-\phi_{\mathbf u}$.
To be compared with the CAMB power spectra, 
$\Delta \lambda_{EE}$ and $\Delta \lambda_{BB}$ should be multiplied with
$l(l+1)/(2\pi)$, where $l=162$ is the multipole our assumed interferometer is most sensitive over. As shown in Fig. \ref{error_B} that $\Delta \lambda_{BB}$ is minimum at $\psi=\phi_{\mathbf u}+\pi/4$ in $V_{Q^\prime}$ measurement and at $\psi=\phi_{\mathbf u}$ in $V_{U^\prime}$ measurement.
$\Delta \lambda_{EE}$ and $\Delta \lambda_{BB}$ from $V_{RL}$ or $V_{LR}$ are invariant under the rotation of the polarizer frame.
For the same assumed noise variance, the minimum $\Delta \lambda_{BB}$ from $V_{Q^\prime}$ or $V_{U^\prime}$ is 26 per cent of $\Delta \lambda_{BB}$ from $V_{RL}$ or $V_{LR}$.
It is shown in Fig. \ref{error_E} that $\Delta \lambda_{EE}$ is minimum at $\psi=\phi_{\mathbf u}$ in $V_{Q^\prime}$ measurement and at $\psi=\phi_{\mathbf u}+\pi/4$ in $V_{U^\prime}$ measurement. The minimum $\Delta \lambda_{EE}$ from $V_{Q^\prime}$ or $V_{U^\prime}$ is 1.2 times bigger than
the $\Delta \lambda_{EE}$ from $V_{RL}$ or $V_{LR}$.
$V_{RL}$ and $V_{RL}$ have information on both of $Q^\prime$ and $U^\prime$, while $V_{Q^\prime}$ and $V_{U^\prime}$ have information on either of $Q^\prime$ and $U^\prime$.
It may seem odd that $\Delta \lambda_{BB}$ of $V_{Q^\prime}$ and $V_{U^\prime}$ can be smaller than $\Delta \lambda_{BB}$ of $V_{RL}$ and $V_{RL}$, though more informations 
are contained in $V_{RL}$ and $V_{RL}$.
$\Delta \lambda_{BB}$ is the estimation error marginalized over $\lambda_{EE}$
, since the likelihood in the simulated observation is the function of two parameters, $\lambda_{EE}$ and $\lambda_{BB}$. 
$\Delta \lambda_{BB}$ is given by $\lambda^0_{BB}-\lambda^1_{BB}$, where 
\[\left.\frac{\partial \mathcal{L}_{B}(\lambda_{BB})}{\partial \lambda_{BB}}\right|_{\lambda_{BB}=\lambda^0_{BB}}=0,\;\;\;\;\;\;\;\;\;\;\;
\frac{\mathcal{L}_{B}(\lambda^1_{BB})}{\mathcal{L}_{B}(\lambda^0_{BB})}
=e^{-1/2}.\]
$\mathcal{L}_{B}(\lambda_{BB})$, which is the likelihood function marginalized over E mode band power, is as follows:  
\begin{eqnarray*}
\mathcal{L}_{B}(\lambda_{BB})=\int d\lambda_{EE}\;\mathcal{L}(\lambda_{EE},\lambda_{BB},\lambda_{EB}=0).
\end{eqnarray*}
$\mathcal{L}(\lambda_{EE},\lambda_{BB})$ becomes less sensitive to the variation of
$\lambda_{EE}$ with reduced contribution of E mode to visibilities.
It makes the $\mathcal{L}_{B}(\lambda_{BB})$ more sharply peaked  around $\lambda^0_{BB}$, which leads to 
the reduction of $\Delta \lambda_{BB}$.

The 1$\sigma$ error on power spectra estimation is the sum of sample variance and noise variance \citep{CMB:strategy}:
$\Delta C_l\sim C_l/(2l+1)+N$, where $N$ is noise variance. 
In the case of circular polarizers, $\Delta_{BB}$ is smaller than $\Delta_{EE}$
due to the smaller sample variance of B mode than that of E mode, even though visibilities with circular polarizers have equal sensitivity to both E and B mode. 

\section{feedhorn array}
\label{horn_array}
For optimization, the polarizer frame should be oriented with certain rotation 
from all the baselines formed by the feedhorn the polarizer is associated with.
When there are just two feedhorns, there is a single baseline and
orienting the polarizer frame to specific rotation physically is trivial. 
But in most of real CMB experiments, an array consisting of more than two feedhorns  are employed.
When all the feedhorns (apertures) are aligned on a straight line, orienting polarizers with specific rotation from all the associated baselines is also trivial even for a (N$>$2) feedhorn array.
When the feedhorn array is configured on some two dimensional pattern for the optimal uv coverage \citep{Guyon:uv_coverage,DASI:I}, it is not possible to realize specific rotation of the polarizer frame from all the associated baselines. 
We can achieve the optimal polarizer rotation for a array of multiple feedhorns (N$>$2)
by forming certain linear combination of visibilities as follows.
With the employment of orthomode transducer (OMT), each baseline measures the real and imaginary part of $V_{Q^\prime}$ and
$V_{U^\prime}$ respectively.
Four measured values are obtained such that $V_1=\mathrm{Re}[V_{Q^\prime}]$, $V_2=\mathrm{Im}[V_{Q^\prime}]$, $V_3=\mathrm{Re}[V_{U^\prime}]$, $V_4=\mathrm{Im}[V_{U^\prime}]$.
With Eq. \ref{V_Q_spherical} and \ref{V_U_spherical}, $V_1$, $V_2$, $V_3$, and $V_4$ are as follows:
\begin{eqnarray}
\label{V_1}
V_{1}&=&f(\nu)\int \mathrm d \Omega A(\mathbf {\hat n}-\hat {\mathbf n}_A)\\
&&\times\mathrm{Re}\,[e^{-i(2\psi-2\Phi(\mathbf {\hat n}))}
(Q(\mathbf {\hat n})+i U(\mathbf {\hat n}))]\,\cos(\,2\pi\mathbf u\cdot \mathbf {\hat n})\nonumber\\
\label{V_2}
V_{2}&=&f(\nu)\int \mathrm d \Omega A(\mathbf {\hat n}-\hat {\mathbf n}_A)\\
&&\times\mathrm{Re}\,[e^{-i(2\psi-2\Phi(\mathbf {\hat n}))}
(Q(\mathbf {\hat n})+i U(\mathbf {\hat n}))]\,\sin(\,2\pi\mathbf u\cdot \mathbf {\hat n}),\nonumber\\
\label{V_3}
V_{3}&=&f(\nu)\int \mathrm d \Omega A(\mathbf {\hat n}-\hat {\mathbf n}_A)\\
&&\times\mathrm{Im}\,[e^{-i(2\psi-2\Phi(\mathbf {\hat n}))}
(Q(\mathbf {\hat n})+i U(\mathbf {\hat n}))]\,\cos(\,2\pi\mathbf u\cdot \mathbf {\hat n}),\nonumber\\
\label{V_4}
V_{4}&=&f(\nu)\int \mathrm d \Omega A(\mathbf {\hat n}-\hat {\mathbf n}_A)\\
&&\times\mathrm{Im}\,[e^{-i(2\psi-2\Phi(\mathbf {\hat n}))}
(Q(\mathbf {\hat n})+i U(\mathbf {\hat n}))]\,\sin(\,2\pi\mathbf u\cdot \mathbf {\hat n}).\nonumber
\end{eqnarray}
With these, we can form two linear combinations $V_{E}$ and $V_{B}$ such that
\begin{eqnarray*}
\lefteqn{V_{E}=}\\
&&\mathrm{Re}[e^{-i2(\psi_{\mathbf{u}}-\psi)}(V_1+i\:V_3)]+i\,\mathrm{Re}[e^{-i2(\psi_{\mathbf{u}}-\psi)}(V_2+i\:V_4)],\\
\lefteqn{V_{B}=}\\
&&\mathrm{Im}[e^{-i2(\psi_{\mathbf{u}}-\psi)}(V_1+i\:V_3)]+i\,\mathrm{Im}[e^{-i2(\psi_{\mathbf{u}}-\psi)}(V_2+i\:V_4)],
\end{eqnarray*}
With Eq. \ref{V_1}, \ref{V_2}, \ref{V_3} and \ref{V_4}, we can easily show that
\begin{eqnarray*}
V_{E}&=&f(\nu)\int\mathrm d \Omega A(\mathbf {\hat n}-\hat {\mathbf n}_A)\\
&&\times \mathrm{Re}[e^{-i2(\psi_{\mathbf{u}}-\Phi(\mathbf {\hat n}))}(Q(\mathbf {\hat n})+i\,U(\mathbf {\hat n}))]e^{i 2\pi \mathbf u \cdot \mathbf {\hat n}}\\
V_{B}&=&f(\nu)\int\mathrm d \Omega A(\mathbf {\hat n}-\hat {\mathbf n}_A)\\
&&\times \mathrm{Im}[e^{-i2(\psi_{\mathbf{u}}-\Phi(\mathbf {\hat n}))}(Q(\mathbf {\hat n})+i\,U(\mathbf {\hat n}))]e^{i 2\pi \mathbf u \cdot \mathbf {\hat n}}.
\end{eqnarray*}
We can identify the linear combination $V_E$ with the $V_{Q^\prime}$ 
and $V_{B}$ with the $V_{U^\prime}$, whose polarizer frame is aligned with the associated baseline. 
Hence, the linear combination $V_{E}$ is optimized for E mode measurement and $V_{B}$ is for B mode measurement.
Linear combination $V_E$ and $V_B$ enable the optimization for a array of multiple feedhorns (N$>$2). 
\section{CONCLUSION}
As shown in this paper, the contribution from E and B mode to $V_{Q^\prime}$ and $V_{U^\prime}$ are variant under the rotation of the polarizer frame.
In appendix \ref{window_function}, we have shown analytically that the diagonal window functions of $V_{Q^\prime}$ and $V_{U^\prime}$ are maximized, when $x$ axis of the polarizer frame is rotated from the associated baseline with certain angles. 
The 1$\sigma$ error on power spectra estimation from $V_{Q^\prime}$ and $V_{U^\prime}$ measurement are also variant under the rotation of the polarizer frame. Huge variation of 1$\sigma$ error are shown in the simulated observation of the configuration similar to the DASI.
The 1$\sigma$ error on B mode power spectrum estimation is minimized, when $x$ axis of polarizers is rotated from the baseline by $45^\circ$ in $V_{Q^\prime}$ measurement  (when $x$ axis of polarizers is aligned with the baseline in $V_{U^\prime}$ measurement).
Simulation shows that minimum 1$\sigma$ error on B mode power spectra estimation in $V_{Q^\prime}$ or $V_{U^\prime}$ is 26 per cent of that from  $V_{RL}$ or $V_{LR}$, though more information are measured in $V_{RL}$ or $V_{LR}$.
The simulation also shows that the E/B mixing in $V_{Q^\prime}$ or $V_{U^\prime}$ can
be as low as 23 per cent of that in $V_{RL}$ or $V_{LR}$.
With choice of polarizer rotation from the baseline, we can achieve B mode sensitivity 
and E/B separability from $V_{Q^\prime}$ or $V_{U^\prime}$ measurement several times better than  $V_{RL}$ or $V_{LR}$ measurement in the same configuration.

For a array of multiple feedhorns (N$>$2), there always exist certain linear combinations
of visibilities, which are equivalent to visibilities of the optimal polarizer rotation. 
B mode optimization can be achieved for a feedhorn array (N$>$2) 
by forming the linear combinations. 
 
With integration time which makes the noise variance equal to the sample variance,
the estimation error on the power spectra is minimized \citep{Bowden:optimization,CMB:strategy}.
In parallel with the choice for the polarizer rotation from the baseline, we can optimize the interferometer for B mode with the integration time which makes the sample variance of B mode equal to the noise variance.
\section{ACKNOWLEDGMENTS}
The author thanks Gregory Tucker, Peter Timbie, Emory Bunn, Andrei Korotkov and Carolina Calderon for useful discussions.
He thanks Douglas Scott for the hospitality during the visit to UBC.
He thanks anonymous referees for thorough reading and helpful comments, which led to significant improvements in the paper.

\appendix

\section{CMB polarization basis vectors and antenna coordinate}
\label{relation}
In all-sky analysis, the CMB polarization at the angular coordinate ($\theta,\phi$) 
are measured in the local reference frame whose axises are 
($\mathbf {\hat e}_{\theta}$, $\mathbf {\hat e}_{\phi}$, $\mathbf {\hat e}_{r}$).
Let's call this coordinate frame `the local CMBP frame' from now on.
Consider the polarization observation of antenna pointing ($\theta_A$,$\phi_A$). 
A global coordinate frame coincides with the antenna coordinate by Euler rotations $\textbf{R}_y(\theta_A)\,\textbf{R}_z(\phi_A)$.
Since a global coordinate frame coincides with the local CMBP frame by Euler rotations 
$\textbf{R}_y(\theta)\,\textbf{R}_z(\phi)$, the Euler Rotations $\textbf{R}_z(\gamma)\,\textbf{R}_y(\beta)\,\textbf{R}_z(\alpha)$
coincides the antenna coordinate frame with the local CMBP frame,
where $\textbf{R}_z(\gamma)\,\textbf{R}_y(\beta)\,\textbf{R}_z(\alpha)
\,\textbf{R}_y(\theta_A)\,\textbf{R}_z(\phi_A)=\textbf{R}_y(\theta)\,\textbf{R}_z(\phi)$.
Therefore, the local CMBP frame is in rotation from the antenna coordinate by the Euler angles $(\alpha,\beta,\gamma)$ as follows:   
\begin{eqnarray*}
\alpha&=&\tan^{-1}\left[\frac{\sin\theta\sin(\phi-\phi_A)}{\sin\theta\cos\theta_A\cos(\phi-\phi_A)-\cos\theta\sin\theta_A}\right],\\
\beta&=&\cos^{-1}\left[\cos\theta\cos\theta_A+\sin\theta\sin\theta_A\cos(\phi-\phi_A)\right],\\
\gamma&=&\tan^{-1}\left[\frac{\sin\theta_A\sin(\phi-\phi_A)}{-\sin\theta\cos\theta_A+\cos\theta\sin\theta_A\cos(\phi-\phi_A)} \right],
\end{eqnarray*}
where the Euler angles $(\alpha,\beta,\gamma)$ can be obtained from
$\textbf{R}_z(\gamma)\,\textbf{R}_y(\beta)\,\textbf{R}_z(\alpha)
=\textbf{R}_y(\theta)\,\textbf{R}_z(\phi)\,\textbf{R}^{-1}_z(\phi_A)\,\textbf{R}^{-1}_y(\theta_A)$.
In most CMB polarization experiments, where polarizers are attached to the other side of feedhorns, incoming rays go through polarizers after feedhorns. After passing through a feedhorn, an incoming off-axis ray becomes an on-axis ray. Then the local CMBP frame of the ray after the feedhorn system is simply in azimuthal rotation 
$\alpha+\gamma$ from the antenna coordinate. 
Therefore, $\Phi$ in Eq. \ref{Phi} is
\begin{eqnarray*}
\Phi&=&\tan^{-1}\left[\frac{\sin\theta\sin(\phi-\phi_A)}{\sin\theta\cos\theta_A\cos(\phi-\phi_A)-\cos\theta\sin\theta_A}\right]\\
&&+\tan^{-1}\left[\frac{\sin\theta_A\sin(\phi-\phi_A)}{-\sin\theta\cos\theta_A+\cos\theta\sin\theta_A\cos(\phi-\phi_A)} \right].
\end{eqnarray*}

\section{window functions}
\label{window_function}
\subsection{flat sky approximation in small angle limit}
As discussed in \S \ref{V_flat_sky}, visibilities with flat sky approximation are as follows: \begin{eqnarray*}
V_{Q^\prime}&=&f(\nu) \int \mathrm d^2 \mathbf u^\prime \tilde A(\mathbf u-\mathbf u^\prime)\\
&&\times\;[\cos(2(\psi-\phi_{\mathbf u^\prime}))\tilde E(\mathbf u^\prime)+\sin(2(\psi-\phi_{\mathbf u^\prime}))\,\tilde B(\mathbf u^\prime)]\nonumber,\\
V_{U^\prime}&=&f(\nu) \int \mathrm d^2 \mathbf u^\prime\tilde A(\mathbf u-\mathbf u^\prime)\\
&&\times\;[-\sin(2(\psi-\phi_{\mathbf u^\prime}))\tilde E(\mathbf u^\prime)+\cos(2(\psi-\phi_{\mathbf u}))\,\tilde B(\mathbf u^\prime)]\nonumber,\\
V_{RL}&=&f(\nu) \int \mathrm d^2 \mathbf u^\prime\tilde A(\mathbf u-\mathbf u^\prime)\\&&\times\left[e^{-i2\,(\psi-\phi_{\mathbf u^\prime})}(\tilde E(\mathbf u^\prime)+i\,\tilde B(\mathbf u^\prime))\right],\\
V_{LR}&=&f(\nu) \int \mathrm d^2 \mathbf u^\prime\tilde A(\mathbf u-\mathbf u^\prime)\\&&\times\left[e^{i2\,(\psi-\phi_{\mathbf u^\prime})}(\tilde E(\mathbf u^\prime)-i\,\tilde B(\mathbf u^\prime))\right].
\end{eqnarray*}
For multipole $\ell>60$, the following relations between the flat sky power spectra
and the exact power spectra from spherical sky works within one percent error  \citep{White:interferometer}:
\begin{eqnarray}
\label{P_E}\langle E(\mathbf u)E^*(\mathbf u^\prime)\rangle  &\approx&\left. C^{EE}_l\,\delta(\mathbf u-\mathbf u^\prime)\right |_{l=2\pi u},\\
\label{P_B}\langle B(\mathbf u)B^*(\mathbf u^\prime)\rangle  &\approx&\left. C^{BB}_l\,\delta(\mathbf u-\mathbf u^\prime)\right |_{l=2\pi u},\\
\langle E(\mathbf u)B^*(\mathbf u^\prime)\rangle&=&0.
\end{eqnarray}
With the correspondence of the power spectra between flat sky and spherical sky, it can be easily derived that diagonal elements of E/B window functions and their derivatives with respect to the rotation of the polarizer frame, $\psi$, are as follows: 
\begin{enumerate}
\item 
$\langle V_{Q^\prime}(\mathbf u) {V_{Q^\prime}(\mathbf u)}^*\rangle$,
\begin{eqnarray*}
\lefteqn{W^{EE}(\mathbf u,u^\prime)=}\\
&&f^2(\nu)\; u^\prime\int \cos^2(2\psi-2\phi_{\mathbf u^\prime})\tilde A(\mathbf u-\mathbf u^\prime)\tilde A(\mathbf u-\mathbf u^\prime)\mathrm d \phi_{\mathbf u^\prime},\\
\lefteqn{W^{BB}(\mathbf u,u^\prime)=}\\
&&f^2(\nu)\; u^\prime\int \sin^2(2\psi-2\phi_{\mathbf u^\prime})\tilde A(\mathbf u-\mathbf u^\prime)\tilde A(\mathbf u-\mathbf u^\prime)\mathrm d \phi_{\mathbf u^\prime}.\\
\end{eqnarray*}

\label{derivative_W_Q_flat}
\begin{eqnarray*}
\frac{\partial W^{EE}(\mathbf u,u^\prime)}{\partial \psi}&=&
f^2(\nu)\;u^\prime\int \mathrm d \phi_{\mathbf u^\prime} \tilde A^2(\mathbf u-\mathbf u^\prime)\\
&&\times [-2\sin 4\psi\cos 4\phi_{\mathbf u^\prime}+2\cos 4\psi\sin 4\phi_{\mathbf u^\prime}],\\
\frac{\partial W^{BB}(\mathbf u,u^\prime)}{\partial \psi}&=&
f^2(\nu)\;u^\prime\int \mathrm d \phi_{\mathbf u^\prime} \tilde A^2(\mathbf u-\mathbf u^\prime)\\
&&\times [2\sin 4\psi\cos 4\phi_{\mathbf u^\prime}-2\cos 4\psi\sin 4\phi_{\mathbf u^\prime}].\\ 
\end{eqnarray*}

\item 
$\langle V_{U^\prime}(\mathbf u) {V_{U^\prime}(\mathbf u)}^*\rangle$,
\begin{eqnarray*}
\lefteqn{W^{EE}(\mathbf u,u^\prime)=}\\
&&f^2(\nu)\; u^\prime\int \sin^2(2\psi-2\phi_{\mathbf u^\prime})\tilde A(\mathbf u-\mathbf u^\prime)\tilde A(\mathbf u-\mathbf u^\prime)\mathrm d \phi_{\mathbf u^\prime},\\
\lefteqn{W^{BB}(\mathbf u,u^\prime)=}\\
&&f^2(\nu)\; u^\prime\int \cos^2(2\psi-2\phi_{\mathbf u^\prime})\tilde A(\mathbf u-\mathbf u^\prime)\tilde A(\mathbf u-\mathbf u^\prime)\mathrm d \phi_{\mathbf u^\prime}.\\
\end{eqnarray*}
\label{derivative_W_U_flat}
\begin{eqnarray*}
\frac{\partial W^{EE}(\mathbf u,u^\prime)}{\partial \psi}&=&f^2(\nu)\;u^\prime\int \mathrm d \phi_{\mathbf u^\prime} \tilde A^2(\mathbf u-\mathbf u^\prime)\\
&&\times [2\sin 4\psi\cos 4\phi_{\mathbf u^\prime}-2\cos 4\psi\sin 4\phi_{\mathbf u^\prime}],\\ 
\frac{\partial W^{BB}(\mathbf u,u^\prime)}{\partial \psi}&=&f^2(\nu)\;u^\prime\int \mathrm d \phi_{\mathbf u^\prime} \tilde A^2(\mathbf u-\mathbf u^\prime)\\
&&\times [-2\sin 4\psi\cos 4\phi_{\mathbf u^\prime}+2\cos 4\psi\sin 4\phi_{\mathbf u^\prime}]\\
\end{eqnarray*}

\item 
$\langle V_{RL}(\mathbf u) {V_{RL}(\mathbf u)}^*\rangle$
and $\langle V_{LR}(\mathbf u) {V_{LR}(\mathbf u)}^*\rangle$,
\begin{eqnarray*}
\lefteqn{W^{EE}(\mathbf u,u^\prime)=W^{BB}(\mathbf u,u^\prime)=}\\
&&f^2(\nu)\; u^\prime\int \tilde A(\mathbf u-\mathbf u^\prime)\tilde A(\mathbf u-\mathbf u^\prime)\mathrm d \phi_{\mathbf u^\prime},
\end{eqnarray*}
\label{derivative_W_RL_flat}
\begin{eqnarray*}
\frac{\partial W^{EE}(\mathbf u,u^\prime)}{\partial \psi}=\frac{\partial W^{BB}(\mathbf u,u^\prime)}{\partial \psi}=0.
\end{eqnarray*}
\end{enumerate}

From \ref{derivative_W_RL_flat}, we can see that the diagonal E and B mode window functions in $V_{RL}$ and $V_{LR}$ measurement are invariant under the rotation of the polarizer frame.
From \ref{derivative_W_Q_flat} and \ref{derivative_W_U_flat}, we can see that the derivatives of the diagonal E/B window functions in $V_{Q^\prime}$ and $V_{U^\prime}$ measurement are zero, when
\begin{eqnarray}
\lefteqn{\sin 4\psi\int \mathrm d \phi_{\mathbf u^\prime} \tilde A^2(\mathbf u-\mathbf u^\prime)\cos 4\phi_{\mathbf u^\prime}}\label{psi_condition_flat}\\
&=&\cos 4\psi
\int \mathrm d \phi_{\mathbf u^\prime} \tilde A^2(\mathbf u-\mathbf u^\prime)
\sin 4\phi_{\mathbf u^\prime}]\nonumber.
\end{eqnarray}
$\psi$, which satisfies Eq. \ref{psi_condition_flat}, is  
\begin{eqnarray}
\label{psi_flat}\psi=\frac{1}{4}\tan^{-1}\left(\frac{\int \mathrm d \phi_{\mathbf u^\prime}\sin4\phi_{\mathbf u^\prime}\tilde A^2(\mathbf u-\mathbf u^\prime)}{\int \mathrm d \phi_{\mathbf u^\prime}\cos4\phi_{\mathbf u^\prime}\tilde A^2(\mathbf u-\mathbf u^\prime)}\right).
\end{eqnarray}
A Gaussian beam is a good approximation to many CMB experiments and the Fourier transform of a Gaussian beam is 
\begin{eqnarray*}
\tilde A(\mathbf u-\mathbf u^\prime)&=&\exp[-\frac{|\mathbf u-\mathbf u^\prime|^2\sigma^2}{2}]\\
&=&\exp[-\frac{[u^2+{u^\prime}^2-2uu^\prime\cos(\phi_{\mathbf u^\prime}-\phi_{\mathbf u})]\sigma^2}{2}],
\end{eqnarray*}
where $\sigma=0.4245\;\mathrm{FWHM}$.
With a Gaussian beam, the argument of $\tan^{-1}$ in Eq. \ref{psi_flat} is 
\begin{eqnarray}
\lefteqn{\frac{\int^{2\pi}_{0} \mathrm d \phi_{\mathbf u^\prime}\sin4\phi_{\mathbf u^\prime}\exp[2uu^\prime\cos(\phi_{\mathbf u^\prime}-\phi_{\mathbf u})\sigma^2]}{\int^{2\pi}_{0} \mathrm d \phi_{\mathbf u^\prime}\cos4\phi_{\mathbf u^\prime}\exp[2uu^\prime\cos(\phi_{\mathbf u^\prime}-\phi_{\mathbf u})\sigma^2]}}\nonumber\\&=&\frac{\int^{2\pi-\phi_{\mathbf u}}_{-\phi_{\mathbf u}} \mathrm d \phi\sin4(\phi+\phi_{\mathbf u})\exp[2uu^\prime\cos(\phi)\sigma^2]}{\int^{2\pi-\phi_{\mathbf u}}_{-\phi_{\mathbf u}} \mathrm d \phi\cos4(\phi+\phi_{\mathbf u})\exp[2uu^\prime\cos(\phi)\sigma^2]}\nonumber\\
&=&\frac{\int^{2\pi}_{0} \mathrm d \phi\sin4(\phi+\phi_{\mathbf u})\exp[2uu^\prime\cos(\phi)\sigma^2]}{\int^{2\pi}_{0} \mathrm d \phi\cos4(\phi+\phi_{\mathbf u})\exp[2uu^\prime\cos(\phi)\sigma^2]}\nonumber\\
&=&\frac{\sin4\phi_{\mathbf u}\int^{2\pi}_{0} \mathrm d \phi \cos4\phi\exp[2uu^\prime\cos(\phi)\sigma^2]}{\cos4\phi_{\mathbf u}\int^{2\pi}_{0} \mathrm d \phi\cos4\phi\exp[2uu^\prime\cos(\phi)\sigma^2]}\nonumber\\
&=&\tan4\phi_{\mathbf u}\label{argument}
\end{eqnarray}
From the third line to the fourth line in Eq. \ref{argument}, 
$\int^{2\pi}_{0} \mathrm d \phi\sin4\phi\exp[2uu'\cos(\phi)\sigma^2]=0$ was used.
By plugging Eq. \ref{argument} into Eq. \ref{psi_flat}, we get
\begin{eqnarray*}
\psi&=&\frac{1}{4}\tan^{-1}\left(\tan4\phi_{\mathbf u}\right)\\
&=&\phi_{\mathbf u}+\frac{n\pi}{4}
\;\;\;\;\;\;\;\;\;\;\;\;\;\;\;\;\;(n=\ldots, -2, -1, 0, 1, 2, \ldots) 
,\end{eqnarray*}
where $\phi_{\mathbf u}$ is the orientation of the baseline. 
The diagonal element of E and B mode window functions of $V_{Q^\prime}$ and $V_{U^\prime}$
is maximized or minimized, when $x$ axis of the polarizer is in rotation from the baseline
by $-90^\circ$, $-45^\circ$, $0^\circ$, $45^\circ$, $90^\circ$. Since the second derivative of diagonal element of E mode window function has the opposite sign of that of B mode window function,
the diagonal element of B mode window function is minimized at the polarizer rotation which maximizes the diagonal element of E mode window function, and vice versa.
\subsection{spherical sky}
\label{Window_spherical}
Visibilities from spherical sky are as follows:
\begin{eqnarray}
\label{V_Q_EB_spherical_appendix}
V_{Q^\prime}&=&-\frac{1}{2}f(\nu)\int \mathrm d \Omega A(\mathbf {\hat n},\hat {\mathbf n}_A)\,e^{i(2\pi\mathbf u\cdot \mathbf {\hat n})}\\ &&\times[e^{-i(2\psi-2\Phi(\mathbf {\hat n}))}\,(a_{E,lm}+i\,a_{B,lm})\;{}_2Y_{lm}\nonumber\\
&&\;\;\;+e^{i(2\psi-2\Phi(\mathbf {\hat n}))}\,(a_{E,lm}-i\,a_{B,lm})\;{}_{-2}Y_{lm}]\nonumber\\
\label{V_U_EB_spherical_appendix}
V_{U^\prime}&=&\frac{i}{2}f(\nu)\int \mathrm d \Omega A(\mathbf {\hat n},\hat {\mathbf n}_A)\,e^{i(2\pi\mathbf u\cdot \mathbf {\hat n})}\\ &&\times[e^{-i(2\psi-2\Phi(\mathbf {\hat n}))}\,(a_{E,lm}+i\,a_{B,lm})\;{}_2Y_{lm}\nonumber\\
&&\;\;\;-e^{i(2\psi-2\Phi(\mathbf {\hat n}))}\,(a_{E,lm}-i\,a_{B,lm})\;{}_{-2}Y_{lm}]\nonumber\\
\label{V_RL_EB_spherical_appendix}
V_{RL}&=&-f(\nu)\int \mathrm d \Omega A(\mathbf {\hat n},\hat {\mathbf n}_A)\\ 
&&\times (a_{E,lm}+i\,a_{B,lm})\;{}_2Y_{lm}e^{i(2\pi\mathbf u\cdot \mathbf {\hat n}-2\psi+2\Phi(\mathbf {\hat n}))},\nonumber\\
\label{V_LR_EB_spherical_appendix}
V_{LR}&=&-f(\nu)\int \mathrm d \Omega A(\mathbf {\hat n},\hat {\mathbf n}_A)\\
&&\times (a_{E,lm}-i\,a_{B,lm})\;{}_{-2}Y_{lm}e^{i(2\pi\mathbf u\cdot \mathbf {\hat n}+2\psi-2\Phi(\mathbf {\hat n}))}.\nonumber
\end{eqnarray}
With Eq. \ref{V_Q_EB_spherical_appendix}, \ref{V_U_EB_spherical_appendix}, \ref{V_RL_EB_spherical_appendix}, and
\ref{V_LR_EB_spherical_appendix}, visibilities from spherical sky can be expressed as follows:
\begin{eqnarray}
\label{V_Q_RL_lm} 
V_{Q^\prime}(\hat {\mathbf n},\mathbf u)&=&\sum_{l,m}(e^{-i\,2\psi} R_{lm}+e^{i\,2\psi} L_{lm}) a_{E,lm}\nonumber\\
&&+i\,\left(e^{-i\,2\psi} R_{lm}-e^{i\,2\psi} L_{lm}\right)a_{B,lm},\\
\label{V_U_RL_lm} 
V_{U^\prime}(\hat {\mathbf n},\mathbf u)&=&\sum_{l,m}-i(e^{-i\,2\psi} R_{lm}-e^{i\,2\psi} L_{lm}) a_{E,lm}\nonumber\\
&&+\left(e^{-i\,2\psi} R_{lm}+e^{i\,2\psi} L_{lm}\right)a_{B,lm},\\
\label{V_RL_lm} 
V_{RL}(\hat {\mathbf n},\mathbf u)&=&\sum_{l,m}2(a_{E,lm}+i\,a_{B,lm})e^{-i\,2\psi}R_{lm},\\
\label{V_LR_lm} 
V_{LR}(\hat {\mathbf n},\mathbf u)&=&\sum_{l,m}2(a_{E,lm}+i\,a_{B,lm})e^{i\,2\psi}L_{lm},
\end{eqnarray}
where
\begin{eqnarray*}
R_{lm}&=&-\frac{f(\nu)}{2}\int \mathrm d \Omega A(\mathbf {\hat n},\hat {\mathbf n}_A)\;{}_2Y_{lm}e^{i(2\pi\mathbf u_i\cdot \mathbf {\hat n}+2\Phi(\mathbf {\hat n}))},\\
L_{lm}&=&-\frac{f(\nu)}{2}\int \mathrm d \Omega A(\mathbf {\hat n},\hat {\mathbf n}_A)\;{}_{-2}Y_{lm}e^{i(2\pi\mathbf u_i\cdot \mathbf {\hat n}-2\Phi(\mathbf {\hat n}))},\\
\end{eqnarray*}
and $\hat {\mathbf n}_A$ indicates the direction of antenna pointing.
 
Spin $\pm 2$ spherical harmonics have the following form \citep{Zaldarriaga:Polarization_Exp}.
\begin{eqnarray*}
_2Y_{lm}(\mathbf {\hat n}&=&\sqrt{\frac{2l+1}{4\pi}}\left(F_{1,lm}(\theta)+F_{2,lm}(\theta)\right)e^{im\phi},\\
_{-2}Y_{lm}(\mathbf {\hat n}&=&\sqrt{\frac{2l+1}{4\pi}}\left(F_{1,lm}(\theta)-F_{2,lm}(\theta)\right)e^{im\phi},
\end{eqnarray*}
where
$F_{1,lm}$ and $F_{2,lm}$ can be computed in terms of Legendre functions as follows \citep{Kamionkowski:Flm}:
\begin{eqnarray*}
F_{1,lm}(\theta)&=&2\sqrt{\frac{(l-2)!(l-m)!}{(l+2)!(l+m)!}}[(l+m)\frac{\cos\theta}{\sin^2\theta}P^m_{l-1}(\cos\theta)\\
&&-(\frac{l-m^2}{\sin^2\theta}+\frac{1}{2}l(l-1))P^m_l(\cos\theta)],\\
F_{2,lm}(\theta)&=&2\sqrt{\frac{(l-2)!(l-m)!}{(l+2)!(l+m)!}}\frac{m}{\sin^2\theta}[(l+m)P^m_{l-1}(\cos\theta)\\
&&-(l-1)\cos\theta P^m_l(\cos\theta)].\\
\end{eqnarray*}
The covariance properties of the E and B mode are given by 
\begin{eqnarray*}
\langle a_{E,lm} a_{E,l'm'}^* \rangle&=&C^{EE}_{l} \delta_{ll'}\delta_{mm'},\\
\langle a_{B,lm} a_{B,l'm'}^* \rangle&=&C^{BB}_{l} \delta_{ll'}\delta_{mm'},\\
\langle a_{E,lm} a_{B,l'm'}^* \rangle&=&0.\\
\end{eqnarray*}
With these covariance properties, Eq. \ref{V_Q_RL_lm}, \ref{V_U_RL_lm}, \ref{V_RL_lm} and  \ref{V_LR_lm}, it can be easily shown that diagonal elements of E/B window functions and their derivatives with respect to the rotation of the polarizer frame, $\psi$, are as follows: 
\begin{enumerate}
\item \label{W_Q_spherical}
$\langle V_{Q^\prime}(\mathbf u_i) {V_{Q^\prime}(\mathbf u_j)}^*\rangle$,
\begin{eqnarray*}
W^{EE}_{l}&=&\sum_{m}R_{lm}R_{j,lm}^*+L_{lm}L_{j,lm}^*\\
&&+e^{i 4\psi}L_{i,lm}R_{lm}^*+e^{-i 4\psi}R_{lm}L_{j,lm}^*,\nonumber\\
W^{BB}_{l}&=&\sum_{m}R_{lm}R_{lm}^*+L_{lm}L_{lm}^*\\
&&-e^{i 4\psi}L_{lm}R_{lm}^*-e^{-i 4\psi}R_{lm}L_{lm}^*.\nonumber
\end{eqnarray*}
\label{derivative_W_Q_spherical}
\begin{eqnarray*}
\frac{\partial W^{EE}}{\partial \psi}&=&\sum_{m}i 4 e^{i 4\psi}L_{lm}R_{lm}^*-i 4 e^{-i 4\psi}R_{lm}L_{lm}^*,\\
\frac{\partial W^{BB}_{l}}{\partial \psi}&=&\sum_{m}-i 4 e^{i 4\psi}L_{lm}R_{lm}^*+i 4 e^{-i 4\psi}R_{lm}L_{lm}^*.
\end{eqnarray*}
\item \label{W_U_spherical}
$\langle V_{U^\prime}(\mathbf u) {V_{U^\prime}(\mathbf u)}^*\rangle$,
\begin{eqnarray*}
W^{EE}_{l}&=&\sum_{m}R_{lm}R_{lm}^*+L_{lm}L_{lm}^*\\
&&-e^{i 4\psi}L_{lm}R_{lm}^*-e^{-i 4\psi}R_{lm}L_{lm}^*,\nonumber\\
W^{BB}_{l}&=&\sum_{m}R_{lm}R_{lm}^*+L_{lm}L_{lm}^*\\
&&+e^{i 4\psi}L_{lm}R_{lm}^*+e^{-i 4\psi}R_{lm}L_{lm}^*.\nonumber
\end{eqnarray*}
\label{derivative_W_U_spherical}
\begin{eqnarray*}
\frac{\partial W^{EE}_{l}}{\partial \psi}&=&\sum_{m}-i 4 e^{i 4\psi}L_{lm}R_{lm}^*+i 4 e^{-i 4\psi}R_{lm}L_{lm}^*,\\
\frac{\partial W^{BB}_{l}}{\partial \psi}&=&\sum_{m}i 4 e^{i 4\psi}L_{lm}R_{lm}^*-i 4 e^{-i 4\psi}R_{lm}L_{lm}^*.\\
\end{eqnarray*}
\item \label{W_RL_spherical}
$\langle V_{RL}(\mathbf u) {V_{RL}(\mathbf u)}^*\rangle$,
\begin{eqnarray*}
W^{EE}_{l}=W^{BB}_{l}=\sum_{m}\,4 R_{lm}R_{lm}^*\\
\end{eqnarray*}
\begin{eqnarray*}
\frac{\partial W^{EE}_{l}}{\partial \psi}=\frac{\partial W^{BB}_{l}}{\partial \psi}=0.\\
\end{eqnarray*}

\item \label{W_LR_spherical}
$\langle V_{LR}(\mathbf u) {V_{LR}(\mathbf u)}^*\rangle$,
\begin{eqnarray*}
W^{EE}_{l}=W^{BB}_{l}=\sum_{m}\,4 L_{lm}L_{lm}^*.
\end{eqnarray*}
\label{derivative_W_LR_spherical}
\begin{eqnarray*}
\frac{\partial W^{EE}_{l}}{\partial \psi}=\frac{\partial W^{BB}_{l}}{\partial \psi}=0.\\
\end{eqnarray*}
\end{enumerate}

From \ref{W_RL_spherical} and \ref{W_LR_spherical}, we can see that the diagonal E and B mode window functions in $V_{RL}$ and $V_{LR}$ measurement are invariant under the rotation of the polarizer frame.
From \ref{W_Q_spherical} and \ref{W_U_spherical}, we can see that the derivatives of the diagonal E and B mode window functions in $V_{Q^\prime}$ and $V_{U^\prime}$ measurement are zero, when
\begin{eqnarray}
e^{i 4\psi}L_{i,lm}R_{i,lm}^*-e^{-i 4\psi}R_{i,lm}L_{i,lm}^*=0\label{psi_condition}.
\end{eqnarray}
Since the left side of Eq. \ref{psi_condition} is
\begin{eqnarray*}
\lefteqn{-i2\mathrm{Im}[e^{-i 4\psi}R_{i,lm}L_{i,lm}^*]}\\
&=&-i2(-\sin4\psi\mathrm{Re}\,[R_{i,lm}{L_{i,lm}}^*]+\cos4\psi\mathrm{Im}\,[R_{i,lm}{L_{i,lm}}^*]),
\end{eqnarray*}
the following $\psi$ satisfies Eq. \ref{psi_condition}:
\begin{eqnarray}
\label{psi_spherical}\psi=\frac{1}{4}\tan^{-1}\frac{\mathrm{Im}\,[R_{i,lm}{L_{i,lm}}^*]}{\mathrm{Re}\,[R_{i,lm}{L_{i,lm}}^*]}.
\end{eqnarray}
The diagonal elements of window function are independent of the choice of the reference coordinate \citep{White:Window}. So we can choose antenna pointing as $z$ axis without loss of generality.
In the reference frame of our choice, $\Phi(\mathbf {\hat n})$ is reduced to azimuthal angle $\phi$. 
In most of CMB interferometer experiments, primary beampattern are azimuthally symmetric and baselines are coplanar.
Then $R_{lm}$ and $L_{lm}$ are 
\begin{eqnarray}
\label{R_lm_symmetric}
R_{i,lm}&=&-\frac{f(\nu)}{2}\int^\pi_{0} \mathrm d(\theta)\sin\theta A(\theta)\int^{2\pi}_0 \mathrm d\phi\nonumber\\&&\times{}_2Y_{lm}(\theta,\phi) e^{i (2\pi u \sin\theta\cos(\phi-\phi_\mathbf u)-2\psi+2\phi)},\nonumber\\
&=&-\frac{f(\nu)}{2}\sqrt{\frac{2l+1}{4\pi}}\nonumber\\
&&\times\int^{\pi}_{0} \mathrm d(\theta) \sin\theta A(\theta)[F_{1,lm}(\theta)+F_{2,lm}(\theta)]\nonumber\\&&\times \int^{2\pi}_0 \mathrm d\phi e^{i (2\pi u \sin\theta\cos(\phi-\phi_\mathbf u)-2\psi+(m+2)\phi)}\nonumber\\
&=&e^{i\frac{\pi}{2}(m+2)} e^{i (m+2)\phi_{\mathbf u}}\,_2\Theta_{lm},
\end{eqnarray}

\begin{eqnarray}
\label{L_lm_symmetric}
L_{i,lm}&=&-\frac{f(\nu)}{2}\int^\pi_{0} \mathrm d(\theta)\sin\theta A(\theta)\int^{2\pi}_0 \mathrm d\phi\nonumber\\&&\times{}_{-2}Y_{lm}(\theta,\phi) e^{i (2\pi u \sin\theta\cos(\phi-\phi_\mathbf u)+2\psi+2\phi)},\nonumber\\
&=&-\frac{f(\nu)}{2}\sqrt{\frac{2l+1}{4\pi}}\nonumber\\
&&\times\int^{\pi}_{0} \mathrm d(\theta) \sin\theta A(\theta)[F_{1,lm}(\theta)-F_{2,lm}(\theta)]\nonumber\\&&\times \int^{2\pi}_0 \mathrm d\phi e^{i (2\pi u \sin\theta\cos(\phi-\phi_\mathbf u)+2\psi+(m-2)\phi)}\nonumber\\
&=&e^{i\frac{\pi}{2}(m-2)}e^{i (m-2)\phi_{\mathbf u}}\,_{-2}\Theta_{lm},
\end{eqnarray}
where 
\begin{eqnarray*}
_2\Theta_{lm}&=&-f(\nu)\sqrt{\frac{(2l+1)\pi}{4}}\int^{\pi}_{0} \mathrm d \theta \sin(\theta)A(\theta)\\&&\times J_{m+2}(2\pi u\sin(\theta) )(F_{1,lm}(\theta)+F_{2,lm}(\theta)),\\
_{-2}\Theta_{lm}&=&-f(\nu)\sqrt{\frac{(2l+1)\pi}{4}}\int^{\pi}_{0} \mathrm d \theta \sin(\theta)A(\theta)\\&&\times J_{m-2}(2\pi u\sin(\theta) )(F_{1,lm}(\theta)-F_{2,lm}(\theta)).\\
\end{eqnarray*}
By plugging Eq. \ref{R_lm_symmetric} and \ref{L_lm_symmetric} into Eq. \ref{psi_spherical}, we get 
\begin{eqnarray*}
\psi&=&\frac{1}{4}\tan^{-1}\frac{\mathrm{Im}\,[R_{i,lm}{L_{i,lm}}^*]}{\mathrm{Re}\,[R_{i,lm}{L_{i,lm}}^*]}\\
&=&\frac{1}{4}\tan^{-1}\frac{\mathrm{Im} 
[e^{i 4\phi_{\mathbf u}}]_{2}\Theta_{lm}\,_{-2}\Theta_{lm}
}{\mathrm{Re} [e^{i 4\phi_{\mathbf u}}]_{2}\Theta_{lm}\,_{-2}\Theta_{lm}}\\
&=&\phi_{\mathbf u}+\frac{n\pi}{4}.
\;\;\;\;\;\;\;\;\;\;\;\;\;\;\;\;\;(n=\ldots, -2, -1, 0, 1, 2, \ldots)
\end{eqnarray*}
$\psi=\phi_{\mathbf u}+\frac{n\pi}{4}$ maximizes or minimizes the diagonal element of E and B mode window functions of $V_{Q^\prime}$ and $V_{U^\prime}$, which is consistent with the result obtained with flat sky approximation.
\label{lastpage}
\end{document}